# Second harmonic generation in a high-Q lithium niobate microresonator fabricated by femtosecond laser micromachining

Jintian Lin,[1,3] Yingxin Xu,[2] Zhiwei Fang,[1,4] Min Wang,[1,3] Jiangxin Song,[1,3] Nengwen Wang,[1] Lingling Qiao,[1] Wei Fang,[2,*] and Ya Cheng[1,†]

[1]*State Key Laboratory of High Field Laser Physics, Shanghai Institute of Optics and Fine Mechanics, Chinese Academy of Sciences, Shanghai 201800, China*

[2]*Department of Optical Engineering, State Key Laboratory of Modern Optical Instrumentation, Zhejiang University, Hangzhou 310027, China*

[3]*University of Chinese Academy of Sciences, Beijing 100049, China*

[4]*School of Physical Science and Technology, ShanghaiTech University, Shanghai 200031, China*

[*]*Electronic mail: wfang08@zju.edu.cn*

[†]*Electronic mail: ya.cheng@siom.ac.cn*

**Abstract:**

We report on fabrication of high Q lithium niobate (LN) whispering-gallery-mode (WGM) microresonators suspended on silica pedestals by femtosecond laser microfabrication. The micrometer-scale (diameter ~82 μm) LN resonator possesses a Q factor of $2.5 \times 10^5$ around 1550 nm wavelength range. Moreover, second harmonic generation with a continuous-wave tunable single-longitudinal-mode pump laser in the on-chip LN microresonator is demonstrated in the on-chip LN microresonator. A fiber taper is employed to couple the pump laser into the microresonator, showing a normalized conversion efficiency of $1.35 \times 10^{-5}$/mW.

In whispering-gallery-mode (WGM) microresonators, the total internal reflection along the smooth circular sidewalls can lead to high quality (Q) factors and small volumes (V) for inducing dramatic enhancement of light fields. Thanks to their excellent properties, WGM microresonators have been frequently employed in various applications such as microlasers [1], sensing [2], optomechanics [3], nonlinear optics [4,5], cavity quantum electrodynamics [6], etc. For nonlinear optical applications, on-chip silica and semiconductor WGM microresonators, which are fabricated based on sophisticated semiconductor lithograph process, have demonstrated pronounced second order ($\chi^{(2)}$ based) and third order ($\chi^{(3)}$ based) nonlinearities [7-10]. However, the semiconductor microresonators frequently suffer from relative high absorption loss due to their small bandgaps, whereas the silica microresonators have a vanishing $\chi^{(2)}$ and a low $\chi^{(3)}$. Thus, development of microresonators based on advanced nonlinear optical materials is highly desirable.

Dielectric crystalline WGM resonators have shown great promise as the next generation nonlinear sources of both classical and nonclassical light, owing to their properties including relatively high nonlinear coefficients and very low intrinsic absorption and large transparent windows [11-13]. In particular, as an important ferroelectric nonlinear crystalline material, lithium niobate (LN) crystal has received significant attention because fast and efficient tuning can be realized on LN microresonators via electro-optic effect [14]. However, due to the limitation of the material growth technique and the lack of efficient fabrication methods, fabrication of high-Q LN resonator usually relies on mechanical polishing approach [15].

Realization of high-Q on-chip sub-millimeter LN resonators remains a challenge [14, 16], which restrains many applications.

In this Letter, we develop a new technique to fabricate high-Q on-chip sub-millimeter LN microdisks based on femtosecond laser micromachining, followed by focused ion beam (FIB) polishing and thermal treatment to reduce boundary scattering. A Q factor of $2.5\times10^5$ has been achieved from a LN microdisk with a diameter of 82 μm. Second harmonic (SH) generation in this microresonator is demonstrated. We have achieved a normalized conversion efficiency of $1.35\times10^{-5}$/mW using a continuous-wave tunable single-longitudinal-mode pump laser.

In this work, commercially available ion-sliced LN thin films bonded by silica on a LN substrate [17] were used for fabricating the free-standing microresonators. The thicknesses of the z-cut single crystal LN thin film and the silica layer were 0.7 μm and 2 μm, respectively. As schematically illustrated in Fig.1, the procedure of fabrication consists of (1) femtosecond laser ablation of the sample which is immersed in water to form a cylindrical post with a total height of ~15 μm [18-19], (2) smoothing the sidewalls of the cylindrical post by FIB milling [20], (3) chemical etching of the sample in a solution of 5% hydrofluoric (HF) diluted with water to form the freestanding LN microdisk on silica pedestal by selectively removing the silica layer under the LN thin film, and (4) high temperature annealing of the sample to reduce the defects generated by FIB.

To form the cylindrical post by femtosecond laser ablation, the femtosecond laser beam (Legend-Elite, Coherent, Inc., center wavelength: ~800 nm, pulse width: ~40 fs, repetition rate: 1 kHz) was focused into the LN thin film sample immersed in water. The sample was fixed on a computer-controlled XYZ translation stage with 1-μm resolution. A variable neutral density filter was used to carefully adjust the average power. An objective lens with a numerical aperture (NA) of 0.80 was used to produce the tightly focused spot with a diameter of ~1 μm. A charged coupled device (CCD) connecting with the computer was installed above the objective lens to monitor the fabrication process in real time. A layer-by-layer annular ablation from the bottom surface to internal substrate with 1 μm interval between the adjacent layers was adopted, so that the ablation always occurred at the interface between the water and the material. In this manner, the ablation debris can be more efficiently removed with the assistance of water. The average power of the laser was chosen to be 0.35 mW for ablation of both the LN thin film and the LN substrate beneath the silica layer, whereas the average laser power was raised to 1 mW for ablation of the silica layer sandwiched between the LN thin film and the bulk LN substrate, because the ablation thresholds of LN crystal and silica glass are different. After the femtosecond laser ablation, a cylindrical post with a total height of ~15 μm (corresponding to a cutting depth of 12 μm into the LN substrate beneath the silica) was produced, as illustrated in Fig. 2(a).

After the laser ablation, a two-step FIB milling was used to smooth the sidewall of the cylindrical post, as shown in Fig. 1(b). Before the FIB milling, the sample was sputter coated with a ~20 nm layer of gold to reduce the charge collection during the FIB milling. The FIB milling was operated twice, beginning with a coarse milling and followed with a fine one. In the coarse milling, a 30-keV ion beam with a beam current of 4 nA was used to polish the sidewall; whereas in the fine milling, the beam current was reduced to 1 nA. The milling was stopped at a depth of 3 μm from the top surface. The total FIB milling process took ~15 min. After the FIB milling, the smoothness of the sidewall is significantly improved, as evidenced in Figs. 2(b) and (c).

To form the freestanding microresonator (i.e., a thin disk sitting on top of a micro-pedestal), the silica layer sandwiched between the LN thin film and the LN substrate needs to be selectively etched to form the micro-pedestal under the LN microdisk. Therefore, the sample was subjected to a ~8 min bath in a solution of 5% HF diluted with water to form the silica pedestal, and the upper LN microdisk served as the microresonator.

At last, a thermal annealing (500 °C for 2 hrs in air) was applied to further reduce the defects generated in the rim of the LN thin disk caused by the FIB process. It should be mentioned that since the top and bottom surfaces of the LN thin film naturally process an ultra-high smoothness with a surface roughness as low as 0.35 nm [17], after smoothing the sidewall and reducing the scattering defects in the rim area, a high Q factor of the LN microresonator can be ensured.

An evanescent fiber taper coupling was employed to measure the Q factors of fabricated LN microresonators. The light emitted from an external cavity continuous wave tunable laser diode with a wavelength around 1550 nm was coupled into the fiber whose central region was tapered down to about 1 μm by heating and stretching a section of a commercial optical fiber (Corning, SMF-28). A piezostage was used to control the relative position between the microresonator and the fiber taper waist to gain an efficient evanescent coupling. The transmitted spectrum measured from the output end of the fiber taper showed a series of sharp dips in the spectrum at the WGM resonant wavelengths. The transmission spectra of the fiber taper coupled to the microresonators before and after the annealing are depicted in Fig. 3 for comparing the Q factors obtained under these two conditions. A coarse scanning over the entire wavelength span was first performed to check the WGM resonances, then a fine scanning around the resonance wavelength was performed to measure the linewidth of the dip inferred from the Lorentzian fit. The resonance at 1554.28 nm wavelength showed a loaded Q factor of $5.2 \times 10^4$ (Fig. 3(b)) for the microresonator with diameter of 55 μm before the annealing. However, after the annealing, the Q factor of the same microresonator was significantly improved to $1.6 \times 10^5$ around the resonance at 1554.90 nm, as evidenced in Fig. 3(c). Moreover, we found that the Q factor increases with the increasing diameter of the fabricated LN microresonator. The Q factor of the microresonator with a diameter of 82 μm was measured to be $2.5 \times 10^5$ after the annealing process.

To examine the nonlinear optical performance of such crystalline WGM microresonators, SH generation was carried out in the visible wavelength range with the microresonator of the diameter of 82 μm. A continuous wave tunable single-longitudinal-mode Ti-sapphire ring laser (Matisse TX-light, Spectra Physics, Inc.) was used as pump source, which had a spectral linewidth of ~60 kHz. The laser was coupled into the fiber taper with a waist of ~ 1 μm, which was in direct contact with (i.e., over coupling condition) the sidewall of the microresonator to increase the efficiency and stability of coupling of the pump light into the microresonator [10]. An online polarization controller was used to adjust the polarization of the pump light. The SH generated in the microresonator was collected and directed into the grating spectrometer (TriVista, Princeton Instruments Inc.) by the same fiber taper. A band-pass optical filter (400±40 nm) and a short-pass filter (<500 nm) were placed before the spectrometer for removing the pump laser. The width of the entrance slit of the spectrometer was set to be ~50 μm. The maximum SH signal was obtained by monitoring the SH intensity while scanning the pump wavelength between 785 and 815 nm with a step size of 0.4 nm. Once the maximum SH signal was detected with such a coarse scan, we carried out another fine scan of the pump wavelength with a step size of 0.01 nm around the pump wavelength at which the maximum SH signal was generated. Figures 4(a) and (b) plot the transmission spectra of the pump light ($\lambda_P$=799.884 nm) and SH signal ($\lambda_s$=399.922 nm), respectively. The side view optical microscope image of the microresonator is shown in Fig. 4(c), where the SH generation is clearly visible (violet) after removing the pump light with the bandpass filter. Figure 4(d) plots the measured SH conversion efficiency ($\eta = P_{SH}/P_f$, where $P_{SH}$ and $P_f$ are the powers of the SH and the pump laser, respectively) as a function of the power of pump laser coupled into the fiber taper. The normalized conversion efficiency, which can be obtained from the slope of the fitting line in Fig. 4(d) (i.e., the red line in Fig. 4(d)), reaches $1.35\times10^{-5}$/mW. The linear fit also indicates that the power of SH is proportional to the square of the pump laser power, i.e., $P_{SH} = P_f^2$.

In conclusion, we demonstrate the fabrication of on-chip LN microresonators on single crystal LN thin film wafer by femtosecond laser 3D micromachining. The highest Q factor is measured to be $2.5\times10^5$ around 1550 nm wavelength range for a LN microresonator of a diameter of 82 μm. We have also demonstrated SH generation with the fabricated microresonator and obtained a normalized conversion efficiency of $1.35\times10^{-5}$/mW. Since our technique relies on high precision ablation of materials with femtosecond laser pulses, it is material insensitive and can be extended for fabricating on-chip high-Q microresonators in other crystalline materials.

The work is supported by National Basic Research Program of China (No. 2014CB921300), NSFC (Nos. 61275205, 11104245, 61108015, 61008011, 11174305, 11104294, and 61205209), and the Fundamental Research Funds for the Central Universities.

**References:**

[1] S. L. McCall, A. F. J. Levi, R. E. Slusher, S. J. Pearton, and R. A. Logan, Appl. Phys. Lett. **60**, 289 (1992).

[2] F. Vollmer and S. Arnold, Nature Method **5**, 591 (2008).

[3] T. J. Kippenberg and K. J. Vahala, Science **321**, 1172 (2008).

[4] S. M. Spillane, T. J. Kippenberg, and K. J. Vahala, Nature **415**, 621 (2002).

[5] T. J. Kippenberg, S. M. Spillane, and K. J. Vahala, Phys. Rev. Lett. **93**, 083904 (2004).

[6] D. J. Alton, N. P. Stern, Takao, H. Lee, E. Ostby, K. J. Vahala, and H. J. Kimble, Nature Phys. **7**, 159 (2011).

[7] P. Del'Haye, A. Schliesser, O. Arcizet, T. Wilken, R. Holzwarth, and T. J. Kippenberg, Nature **450**, 1214 (2007).

[8] T. Carmon and K. J. Vahala, Nature Phys. **3**, 430 (2007).

[9] P. S. Kuo, J. Brevo-Abad, and G. S. Solomon, Nature Commun. **5**, 3109 (2014).

[10] S. Mariani, A. Andronico, A. Lemaître, I. Favero, S. Ducci, and G. Leo, Opt. Lett. **39**, 3062 (2014).

[11] V. S. Ilchenko, A. A. Savchenkov, A. B. Matsko, and L. Maleki, Phys. Rev. Lett. **92**, 043903 (2004).

[12] J. U. Fürst, D. V. Strekalov, D. Elser, A. Aiello, U. L. Andersen, Ch. Marquardt, and G. Leuchs, Phys. Rev. Lett. **106**, 113901 (2011).

[13] M. Förtsch, J. U. Fürst, C. Wittmann, D. Strekalov, A. Aiello, M. V. Chekhova, C. Silberhorn, G. Leuchs, and C. Marquardt, Nature Commun. **4**, 1818 (2013).

[14] A. Guarino, G. Poberaj, D. Rezzonico, R. Degl'Innocenti, and P. Günter, Nature Photon. **1**, 407 (2007).

[15] A. A. Savchenkov, V. S. Ilchenko, A. B. Matsko, and L. Maleki, Phys. Rev. A **70**, 051804(R) (2004).

[16] T.-J. Wang, J.-Y. He, C.-A. Lee, and H. Niu, Opt. Express **20**, 28119 (2012).

[17] G. Poberaj, H. Hu, W. Sohler, and P. Güter, Laser Photon. Rev. **6**, 488 (2012).

[18] Y. Li, K. Itoh, W. Watanabe, K. Yamada, D. Kuroda, J. Nishii, and Y. Jiang, Opt. Lett. **26**, 1912 (2001).

[19] J. Lin, Y. Xu, J. Song, B. Zeng, F. He, H. Xu, K. Sugioka, W. Fang, and Y. Cheng, Opt. Lett. **38**, 1458 (2013).

[20] J. Lin, Y. Xu, J. Tang, N. Wang, J. Song, F. He, W. Fang, and Y. Cheng, Appl. Phys. A (published online, DOI 10.1007/s00339-014-8388-1), 2014.

**Figure Captions:**

Fig. 1. Procedures of fabrication of a LN microresonator by water-assisted femtosecond laser ablation, followed by FIB milling, selective chemical etching, and annealing.

Fig. 2. (a) SEM image of a cylindrical post formed after femtosecond laser ablation; (b) and (c) SEM images of two cylindrical posts with diameters of 55 μm and 33 μm, respectively, after the FIB milling; (d) SEM image (top view) of the 55-μm diameter microresonator after the chemical etching and annealing. Inset in (d): side view of the microresonator, showing the freestanding edge of the LN microresonator.

Fig. 3. (a) Transmission spectrum of the fiber taper coupled with the microresonator (diameter ~55 μm) before annealing, (b) Lorentzian fit (red solid line) of measured spectrum around the resonant wavelength at 1554.28 nm (black dotted line), showing a Q factor of $5.2 \times 10^4$, (c) Lorentzian fit (red solid line) of measured transmission spectrum of the microresonator (diameter ~55 μm) after annealing around the resonant wavelength at 1554.90 nm (black dotted line), showing an improved Q factor of $1.6 \times 10^5$, (d) Lorentzian fit (red solid line) of measured transmission spectrum of the microresonator (diameter ~82 μm) after annealing around the resonant wavelength at 1553.83 nm (black dotted line), showing an improved Q factor of $2.5 \times 10^5$.

Fig. 4. (a) The typical spectrum of pump laser for efficiently generating second harmonic. (b) The spectrum of generated SH signal. (c) Optical microscope side view image of the microresonator under the irradiation of the pump laser, showing the SH beam (the violet light) scattering out from the microresonator. (d) SHG conversion efficiency as a function of pump power. The slope indicates a normalized conversion efficiency of $1.35\times10^{-5}$/mW.

Fig.1

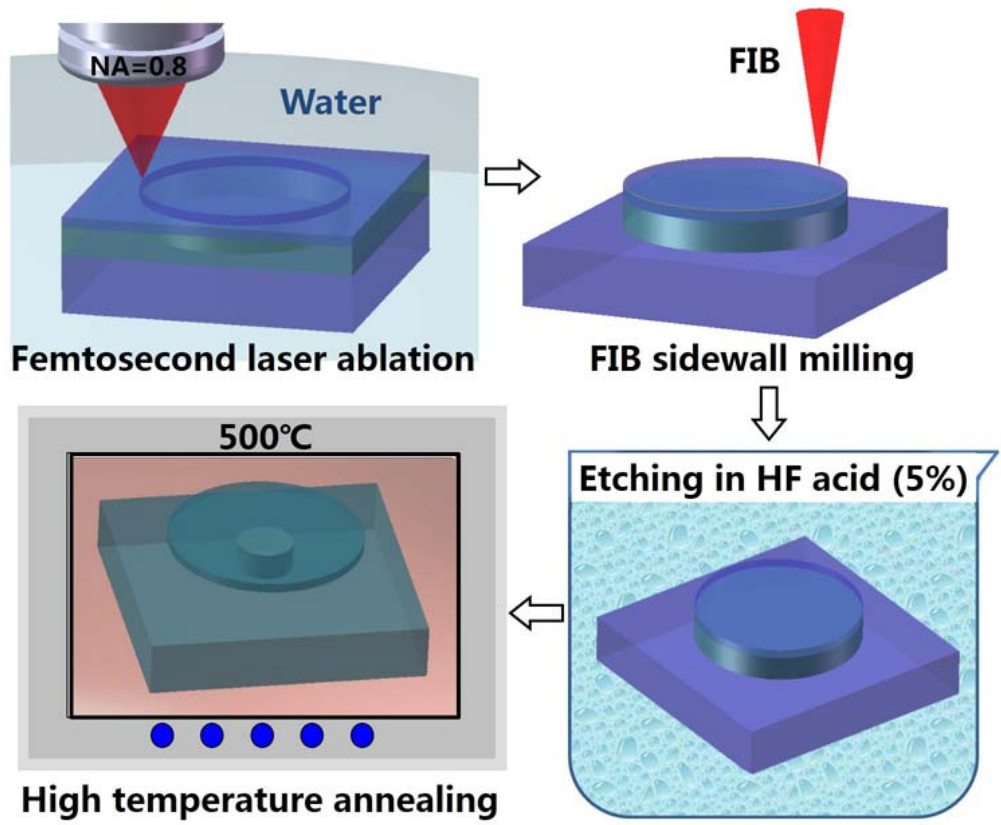

Fig.2

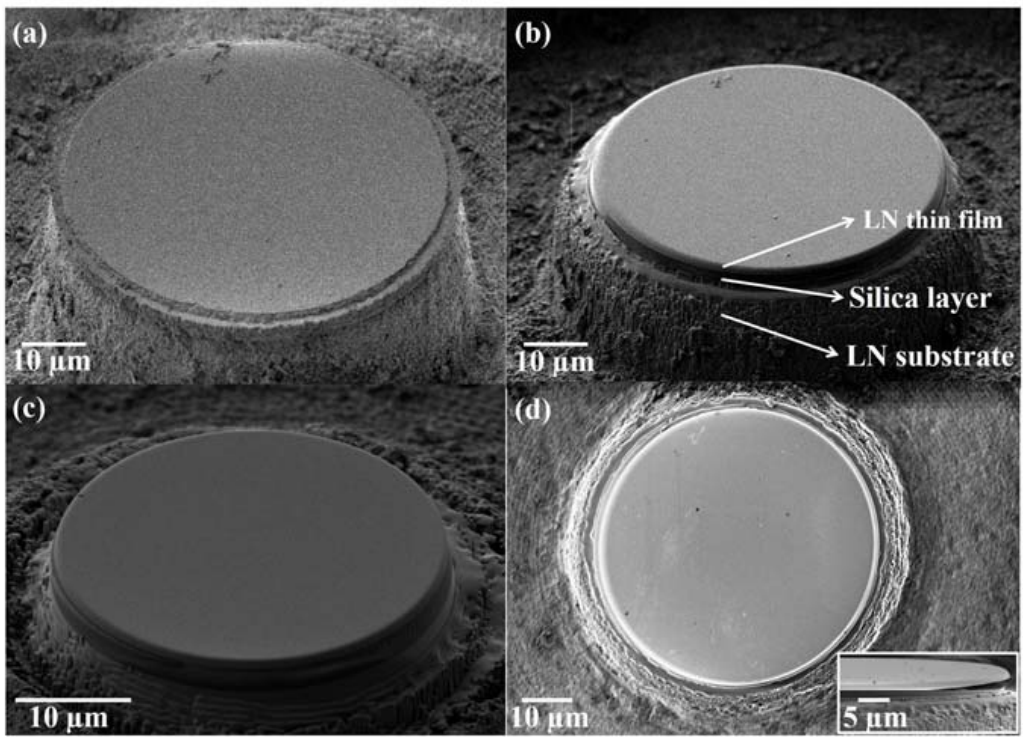

Fig.3

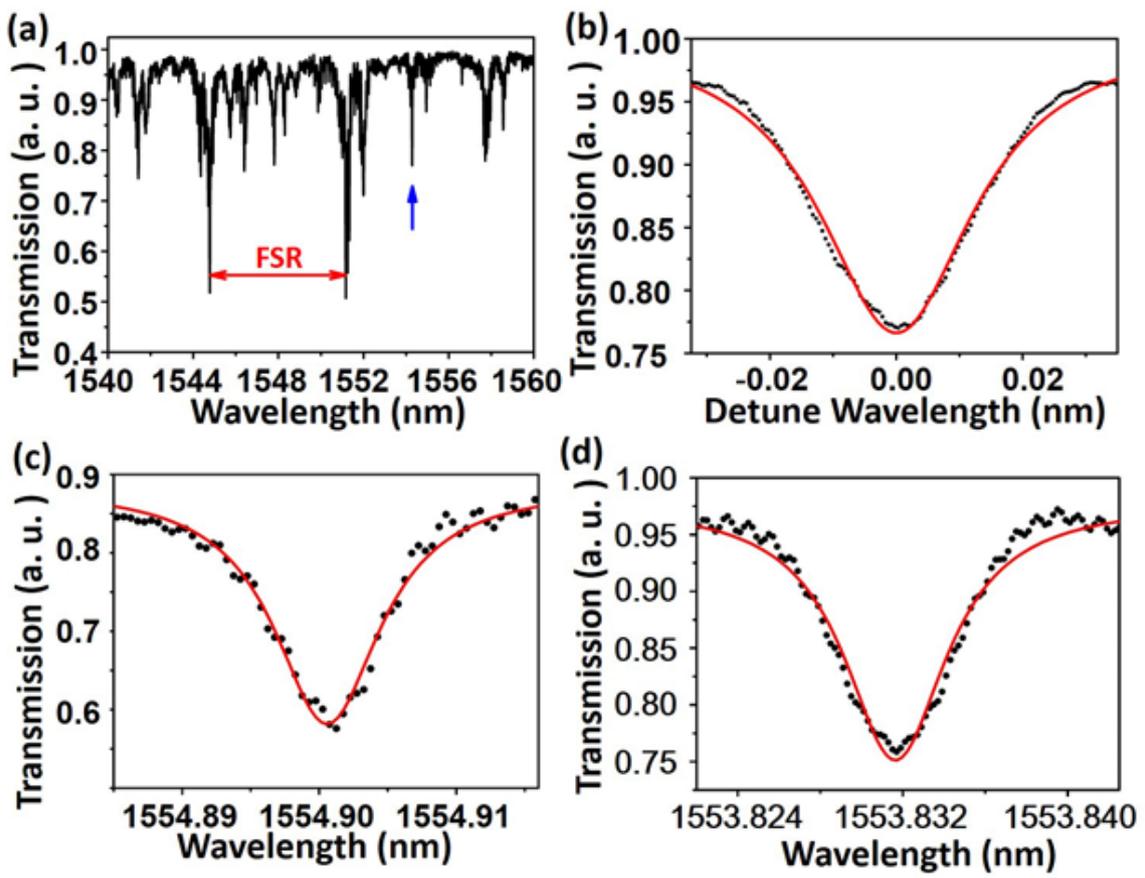

Fig.4

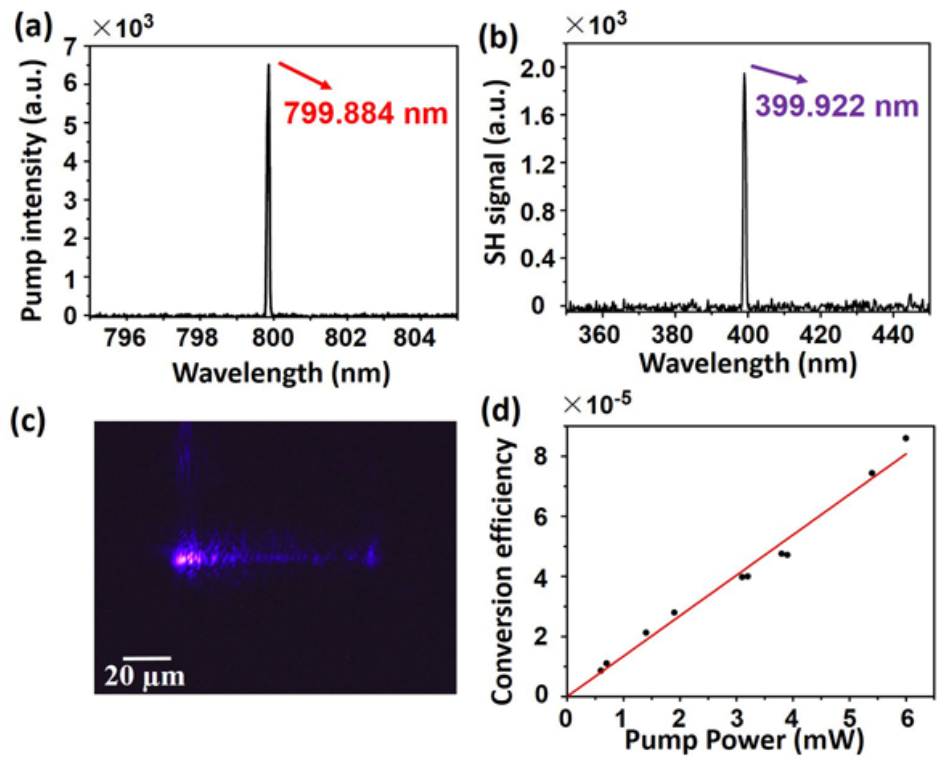